# The Brain Tumor Segmentation (BraTS) Challenge 2023: *Glioma Segmentation in Sub-Saharan Africa Patient Population (BraTS-Africa)*

Maruf Adewole[1,2,†,‡*], Jeffrey D. Rudie[3,†,§*], Anu Gbdamosi[1,4‡], Oluyemisi Toyobo[1,4,‡§], Confidence Raymond[1,†], Dong Zhang[1,†], Olubukola Omidiji[1,5,‡§], Rachel Akinola[6,‡], Mohammad Abba Suwaid[7,‡], Adaobi Emegoakor[8,§], Nancy Ojo[9,§], Kenneth Aguh[10,§], Chinasa Kalaiwo[11,§], Gabriel Babatunde[5,§], Afolabi Ogunleye[6,§], Yewande Gbadamosi[6,§], Kator Iorpagher[12,§], Evan Calabrese[13,14,†,‡], Mariam Aboian[15,†], Marius Linguraru[16,17,†,‡], Jake Albrecht[18,†], Benedikt Wiestler[19,†,‡], Florian Kofler[19,20,†], Anastasia Janas[15,†,§], Dominic LaBella[21,†], Anahita Fathi Kzerooni[22,23,†,‡], Hongwei Bran Li[19,24,25,†], Juan Eugenio Iglesias[24,†], Keyvan Farahani[26,†,‡], James Eddy[18,†], Timothy Bergquist[18,†], Verena Chung[18,†], Russell Takeshi Shinohara[23,27,†], Walter Wiggins[13,†,‡,,], Zachary Reitman[21,†,‡,,], Chunhao Wang[21,†,‡], Xinyang Liu[16,17,†], Zhifan Jiang[16,17,†], Ariana Familiar[22,†], Koen Van Leemput[28,†], Christina Bukas[20,†], Maire Piraud[20,†], Gian-Marco Conte[29,†], Elaine Johansson[30,†], Zeke Meier[31,†,], Bjoern H Menze[19,25,†,‡], Ujjwal Baid[23,32,33,†,‡], Spyridon Bakas[23,32,33,†,‡], Farouk Dako[34,†], Abiodun Fatade[1,4,†], and Udunna C Anazodo[1,35,36,37,†,‡, ¶]

[1] Medical Artificial Intelligence Laboratory (MAI Lab), Lagos, Nigeria
[2] Department of Radiation Biology, Radiotherapy and Radiodiagnosis, University of Lagos, Lagos, Nigeria
[3] Department of Radiology, University of California, San Diego
[4] Crestview Radiology Limited, Lagos, Nigeria
[5] Lagos University Teaching Hospital, Lagos, Nigeria
[6] Lagos State University Teaching Hospital, Ikeja, Lagos, Nigeria
[7] NSIA-Kano Diagnostic Center, Kano Nigeria
[8] Nnamdi Azikiwe University Teaching Hospital, Nnewi, Anambra State, Nigeria
[9] Federal Medical Centre, Abeokuta, Ogun State, Nigeria
[10] Federal Medical Centre, Umahia, Abia State, Nigeria
[11] National Hospital Abuja, FCT, Nigeria
[12] Benue State University Teaching Hospital, Markurdi, Benue State, Nigeria
[13] Duke University Medical Center, Department of Radiology, USA
[14] University of California San Francisco, CA, USA
[15] Yale University, New Haven, CT, USA
[16] Children's National Hospital, Washington DC, USA
[17] George Washington University, Washington DC, USA
[18] Sage Bionetworks, USA
[19] Department of Neuroradiology, Technical University of Munich, Munich, Germany
[20] Helmholtz Research Center, Munich, Germany
[21] Duke University Medical Center, Department of Radiation Oncology, USA
[22] Children's Hospital of Philadelphia, University of Pennsylvania, Philadelphia, PA, USA
[23] Center for AI and Data Science for Integrated Diagnostics (AI2D) & Center for Biomedical Image Computing and Analytics (CBICA), University of Pennsylvania, Philadelphia, PA, USA
[24] Athinoula A Martinos Center for Biomedical Imaging, Massachusetts General Hospital, Boston, MA, USA
[25] University of Zurich, Switzerland
[26] Cancer Imaging Program, National Cancer Institute, National Institutes of Health, Bethesda, MD 20814, USA
[27] Center for Clinical Epidemiology and Biostatistics, University of Pennsylvania, Philadelphia, USA
[28] Department of Applied Mathematics and Computer Science, Technical University of Denmark, Denmark
[29] Mayo Clinic, MN, USA
[30] Precision FDA, U.S. Food and Drug Administration, Silver Spring, MD, USA
[31] Booz Allen Hamilton, McLean, VA, USA
[32] Department of Radiology, Perelman School of Medicine, University of Pennsylvania, Philadelphia, PA, USA
[33] Department of Pathology and Laboratory Medicine, Perelman School of Medicine, University of Pennsylvania, Philadelphia, PA, USA
[34] Center for Global Health, Perelman School of Medicine, University of Pennsylvania, Philadelphia, Pennsylvania, USA
[35] Montreal Neurological Institute, McGill University, Montreal, Canada
[36] Department of Medicine, University of Cape Town, South Africa
[37] Department of Radiation Medicine, University of Cape Town, South Africa

† People involved in the organization of the challenge.
‡ People contributing data from their institutions.
§ People involved in annotation process.
* First Authors
¶ Senior Authors
** Corresponding author: {udunna.anazodo@mcgill.ca}

**Abstract.** Gliomas are the most common type of primary brain tumors. Although gliomas are relatively rare, they are among the deadliest types of cancer, with a survival rate of less than 2 years after diagnosis. Gliomas are challenging to diagnose, hard to treat and inherently resistant to conventional therapy. Years of extensive research to improve diagnosis and treatment of gliomas have decreased mortality rates across the Global North, while chances of survival among individuals in low- and middle-income countries (LMICs) remain unchanged and are significantly worse in Sub-Saharan Africa (SSA) populations. Long-term survival with glioma is associated with the identification of appropriate pathological features on brain MRI and confirmation by histopathology. Since 2012, the Brain Tumor Segmentation (BraTS) Challenge have evaluated state-of-the-art machine learning methods to detect, characterize, and classify gliomas. However, it is unclear if the state- of-the-art methods can be widely implemented in SSA given the extensive use of lower-quality MRI technology, which produces poor image contrast and resolution and more importantly, the propensity for late presentation of disease at advanced stages as well as the unique characteristics of gliomas in SSA (i.e., suspected higher rates of gliomatosis cerebri). Thus, the BraTS-Africa Challenge provides a unique opportunity to include brain MRI glioma cases from SSA in global efforts through the BraTS Challenge to develop and evaluate computer-aided-diagnostic (CAD) methods for the detection and characterization of glioma in resource-limited settings, where the potential for CAD tools to transform healthcare are more likely.

## 1   Introduction

Brain tumors are among the deadliest types of cancer. Approximately 80% of individuals with glioma - the predominant and the most malignant form of primary brain tumor - die within two years of diagnosis[1]. In contrast, over 90% of individuals with prostate or breast cancer are expected to survive after 5 years of diagnosis regardless of the malignant status[2]. Brain tumors in general are challenging to diagnose, hard to treat and inherently resistant to conventional therapy because of the challenges in delivering drugs to the brain. Years of extensive research to improve diagnosis, characterization, and treatment of glioma have decreased mortality rates in the U.S by 7% over the past 30 years[3]. Although modest, these research innovations have not translated to improvements in survival for adults and children in low- and middle-income countries (LMICs), particularly in African populations. The death rates from glioma in sub-Saharan Africa (SSA) are among the highest in the world and continue to rise. While glioma deaths rates dropped in the Global North by 10-30% from 1990 to 2016, the rates in SSA rose on average by ~25%[3]. This disparity could be due to several overlapping factors including, delayed presentation[4,5], high incidence of infectious disease comorbidities such as HIV[6], severe shortage of healthcare infrastructure including therapeutic options[5], and lack of skilled expertise in diagnosis (neuroradiologists and neuropathologists) and treatment (neurosurgeon, neurooncologists, and medical physicists)[7] of glioma. As standard treatment options for glioma (surgical resection, radiation, and concomitant chemotherapy) increasingly become available in SSA, the novel use of machine learning to advance early detection, identify precise treatment targets, and predict progression and treatment response will further widen survival disparities, if SSA populations are not included in innovative solutions that collectively benefit all patients.

 Long-term survival (>3years) with glioma is not only associated with socioeconomic, environmental, and occupational factors, rather, patient survival is largely influenced by molecular, genetic, and clinical factors including tumor volume, tumor grade, age at diagnosis and histologic findings[8,9]. Detection and analysis to measure glioma volume and classify glioma is currently dependent on identifying appropriate pathological features on brain magnetic resonance imaging (MRI) and confirmation by histopathology evaluation of biopsied tissue[10,11]. Computer-aided diagnosis using machine learning (ML) to segment brain tumors holds promise in increasing accuracy of tumor diagnosis, early detection, classification and in predicting tumor recurrence and patient survival [11-13]. ML algorithms for assessment of tumor burden and treatment response based on tumor segmentation on brain MRI were recently shown to outperform human readers in a large multi-center study of glioma patients[13]. More importantly ML can close survival disparity gaps by overcoming challenges in low-resourced setting, where time consuming manual evaluations are limited to the rare centers in urban areas that can afford highly skilled expert personnel to perform tumor analysis.

Since 2012, the Brain Tumor Segmentation (BraTS) Challenge has exploited the powerful role of ML in glioma diagnosis, focusing on evaluation of state-of-the art methods for tumor segmentation, classification and more recently survival prediction[14–16]. Each year BraTS provides open and accessible multiparametric MRI training data, ground-truth annotations and quantitative metrics for benchmarking performance and clinical utility of ML methods for glioma diagnostics. However, it is unclear if the state-of-the art ML methods developed using BraTS data can be widely implemented for clinical use in SSA, particularly given the extensive use of lower quality MRI technology in the region[17, 18]. Brain MRI typically acquired in SSA have poor image contrast and resolution (Figure 1) and may require further advanced image pre-processing to enhance their resolution[19] before application of ML methods for tumor segmentation, classification, or outcome prediction. Therefore, there is a pressing need to provide appropriately curated and annotated brain MRI images from SSA that can represent real world standard of care images acquired in SSA or other low-resourced settings for ML glioma applications[18], especially for development and evaluation of ML approaches that aim to establish critical diagnostic biomarkers through Grand Challenges.

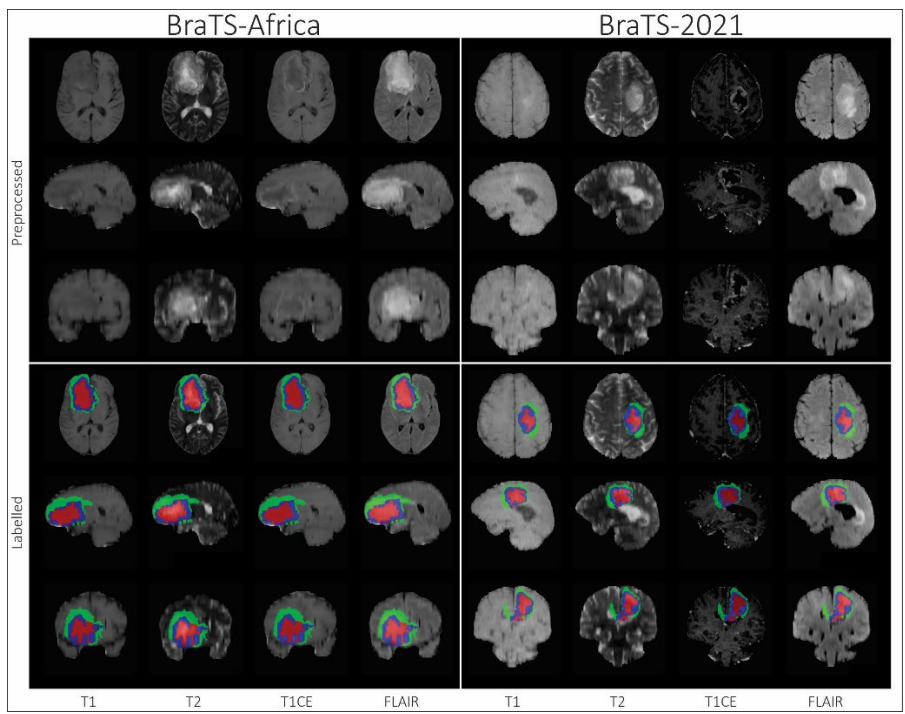

**Figure 1**: **Clinical standard brain MRI typically acquired in Sub-Saharan African populations.** Brain slices of the T1-weighted, T2-weighted, post-gadolinium contrast enhanced T1-weighted (T1CE) and T2 Flair MRI from an SSA glioma patient (left) show the conventional lower image resolution, evident on the sagittal slices compared to the clinical standard images of a representative case from the 2021 BraTS challenge.

## 2    Materials & Methods

### 2.1    Data

This Challenge provides for the first time, annotated training, validation, and testing data sets in adult populations collected through a collaborative network of imaging centres in Africa with support from the Consortium for Advancement of MRI Education and Research in Africa (CAMERA)[1] and funding from the Lacuna Fund[2] in Health Equity. As part of the BraTS 2023 cluster of challenges, this Challenge in SSA populations is a Medical Image and Computing and Computer Assisted Interventions (MICCAI) society registered challenge.

The MICCAI-CAMERA-Lacuna Fund BraTS-Africa 2023 (BraTS-Africa 2023) Challenge dataset is a publicly available retrospective collection of pre-operative glioma data comprising of multi-parametric (mpMRI) routine clinical scans acquired as part of standard clinical care from multiple institutions and different scanners using conventional brain tumor imaging protocols. The differences in imaging systems and variations in clinical imaging protocols result in a vastly heterogeneous image quality, providing real world cases for robust ML methods development. Ground truth annotations of the tumor sub-regions (figure 2) in each case were approved by expert neuroradiologists as described below. The challenge dataset is divided into training (70%), validation (10%), and testing (20%) datasets. The challenge participants are provided training data with the associated ground truth labels and then the validation data without any associated ground truth labels. The testing data and ground truth label are withheld from the challenge participants throughout the duration of the challenge.

The BraTS-Africa challenge is an extension of the BraTS 2023 continuous challenge with the specific task to create ML algorithms to automatically segment intracranial gliomas into three distinct classes using a 3-label system (figure 2), as described in the annotation protocol below. The challenge participants can supplement the BraTS-Africa data with additional public and/or private glioma MRI data (from their own institutions), for the training of their algorithms, provided the supplemental dataset are explicitly and thoroughly described in the methods of submitted manuscripts and u s e d only for scientific publication purposes. Importantly, participants that decide to supplement their training dataset are required to report results using only the BraTS2023 glioma data and results that include the supplemental data and discuss potential result differences.

### 2.1.1  Imaging Data Description

The mpMRI scans for the BraTS-Africa 2023 challenge are image volumes of 1) T1-weighted (T1), 2) post gadolinium (Gd) contrast T1-weighted (T1Gd), 3) T2-weighted (T2), and 4) T2 Fluid Attenuated Inversion Recovery (T2-FLAIR). All imaging data were reviewed by a board-certified radiologist with extensive experience in the field of neuro-oncology.

Standardized pre-processing using the BraTS pre-processing workflow were used to generate the training, validation, and testing dataset[14-16, 20-21]. T h e preprocessing pipeline applied to the BraTS-Africa 2023 challenge data is identical with the pipeline applied on BraTS 2017-2022 challenges. Specifically, the applied pre-processing routines include conversion of the DICOM files to the NIfTI file format, co-registration to the same anatomical template (SRI24)[22], resampling to a uniform isotropic resolution ($1mm^3$), and finally skull-stripping. The pre-processing pipeline is publicly available through the Cancer Imaging Phenomics Toolkit (CaPTk)[3] [21,23,24] and Federated Tumor Segmentation (FeTS) tool[4]. Conversion to NIfTI strips the accompanying metadata from the DICOM images, and essentially removes all Protected Health Information (PHI) from the DICOM headers. Furthermore, skull-stripping mitigates potential facial reconstruction/recognition of the patient[25,26]. The specific approach we have used for skull stripping is based on a novel DL approach that accounts for the brain shape prior and is agnostic to the MRI sequence input[26]. All imaging volumes were then segmented using the STAPLE[27] fusion of previous top-ranked BraTS algorithms, namely, nnU-Net[28], DeepScan[29], and DeepMedic[30] for

---

generation of ground truth labels.

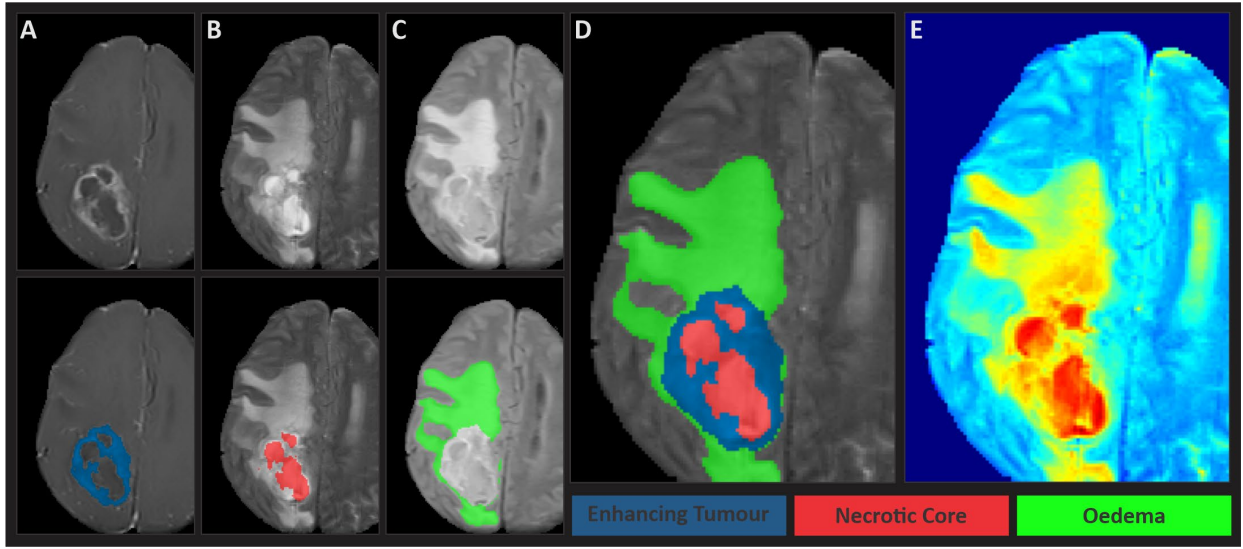

Figure 2: **Glioma sub-regions considered in the MICCAI-CAMERA-LACUNA Fund BraTS 2023 challenge.** Image panels denoting the tumor sub-regions annotated in the different mpMRI scans of a representative case. The image panels A-C denote the regions considered for the performance evaluation of the participating algorithms (top) and three labels overlaid on MRI image highlighted (bottom, D in grayscale and E in RBG color model) (from left to right): panel A) the enhancing tumor (ET - yellow), panel B) the non-enhancing necrotic tumor core (NETC – red), and panel C) the combined surrounding sub-region (SNET - blue).

### 2.1.2    Tumor Annotation Protocol

For each case, the BraTS tumor annotation protocol was used to generate the ground truth labels, to ensure consistency in the ground truth delineations across various annotators. To facilitate the annotation process for BraTS 2021, initial automated segmentations were generated from an nnU-Net[28] model trained on the BraTS 2021 dataset[14-16]. All these segmentation methods and the exact pipeline used to generate the fused automated segmentation has been made publicly available through the Federated Tumor Segmentation (FeTS) platform[5] [31]. The segmented images were refined manually by volunteer trained radiology experts of varying rank and experience, following consistently the BraTS annotation protocol. The volunteer radiology expert annotators were provided with the four processed mpMRI scans (T1, T1Gd, T2, and T2-Flair as well as a T1Gd-T1 subtraction image) along with the fused automated segmentation volume to initiate the manual refinements. The ITK-SNAP[32] software was used for making these refinements. Once the automated segmentations were refined by the annotators, two senior attending board-certified radiologists with 5 or more years of experience, reviewed the segmentations. Depending upon correctness, these segmentations were either approved or returned to the individual annotator for further refinements. This process was followed iteratively until the approvers found the refined tumor sub-region segmentations acceptable for public release and the challenge conduction. The manually refined annotations were finally approved by experienced board-certified attending neuro-radiologists, with more than 5 o r m o r e  years of experience interpreting glioma brain MRI.

    We note that radiologic definition of tumor boundaries, especially in such infiltrative

---

tumors as gliomas, is a well-known problem. In an attempt to offer a standardized approach to assess and evaluate various tumor sub-regions, the BraTS initiative, after consultation with internationally recognized expert neuroradiologists, defined three tumor sub-regions. However, we note that other criteria for delineation could be set, resulting in slightly different tumor sub-regions. For the BraTS-Africa 2023 challenge the sub-regions considered are:

1. **The enhancing tumor (ET)**: includes all portions of the tumor that have noticeable increase in T1 signal on postcontrast images relative to pre-contrast images. This does not include adjacent blood vessels, even if abnormally enlarged. Intrinsic T1 hyperintensity is also NOT included in this label.

2. **Non-enhancing tumor core (NETC)**: includes all portions of the tumor core (i.e. the part that would normally be resected by a surgeon) that do NOT enhance. This includes necrosis, cystic change, calcification, and even exophytic hyperostosis extending into the tumor. This also includes intrinsic T1 hyperintensity such as intratumoral hemorrhage and fat.

3. **Surrounding non-enhancing** flair hyperintensity (SNFH): is the entire extent of FLAIR signal abnormality surrounding the tumor that is not part of the tumor core. For meningiomas, this is often described as "vasogenic edema". This does NOT include non-tumor related FLAIR signal abnormality such as prior infarcts or microvascular ischemic white matter changes.

The BraTS tumor sub-regions (visual features) are image-based and do not reflect strict biologic entities. For example, the ET regions may be defined as hyper-intense signal on T1Gd images. However, in high grade tumors, non-necrotic, non-cystic regions are present that do not enhance and can be separable from the surrounding vasogenic edema, representing non-enhancing infiltrative tumor. Another issue is defining the tumor center in low grade gliomas as it is difficult to differentiate tumor from vasogenic edema, particularly in the absence of enhancement.

### 2.1.3    Common errors of automated segmentations

Building upon observations during all previous BraTS instances, we note some common errors in the automated segmentations. The most typical such errors observed are:

1. Extension of ET into vessels and choroid plexus
2. Under and over segmented areas of SNFH
3. Incorrect segmentation of areas of hemorrhage as ET

### 2.1.4    Performance Evaluation

A baseline approach implemented in a modular open-source framework, namely the Generally Nuanced Deep Learning Framework (GaNDLF)[33] maintained by the MLCommons organization[6]. GaNDLF offers some network architectures, but also allows the user to leverage the functionality of other libraries, such as PILLOW and MONAI. The participants can decide to use GaNDLF to develop their approach or use their own custom source code. For submission, participants are required to package their developed approach in an MLCube container following instructions provided in the Synapse platform. Cube containers are automatically generated by GaNDLF and will be used to evaluate all submissions through the MedPerf platform[34] on each contributing site's data.

---

[6] *https://mlcommons.org/en/*

The evaluation metrics, used in this challenge are the same metrics from previous BraTS challenges:

1. the Dice Similarity Coefficient (DSC), which is commonly used in the assessment of segmentation.

2. the 95% Hausdorff distance (HD) as opposed to standard HD, in order to avoid outliers having too much weight, sensitivity and Specificity to determine whether an algorithm has the tendency to over- or under segment.

3. precision to complement the metric of Sensitivity (also known as recall).

Participants submitted algorithms will be ranked based on the generated metric results on the test cases by computing the summation of their ranks across the average of the metrics described above as a univariate overall summary measure. This measure will decide the overall ranking for each specific team's submission. To visualize the results in an intuitive fashion, the outcome will be plotted via an augmented version of radar plot[35]. Missing results on test cases or if an algorithm fails to produce a result metric for a specific test case, the metric will be set to its worst possible value (0 for the DSC and the image diagonal for the HD). The ranking scheme from previous challenges will be used and in keeping with discussions with the biostatistician involved in the design of The BraTS challenge (Dr Shinohara), and also while considering transparency and fairness to participants. Similar to BraTS 2017-2022, uncertainties in rankings will be assessed using permutational analyses[36]. Performance for the segmentation task will be assessed based on relative performance of each team on each tumor tissue class and for each segmentation measure. These will be combined by averaging ranks for the measures, and statistical significance will be evaluated only for the segmentation performance measures and will be quantified by permuting the relative ranks for each segmentation measure and tissue class per subject of the testing data. This permutation testing would reflect differences in performance that exceeded those that might be expected by chance.

## Participation

The challenge will commence with the release of the training dataset, which will consist of imaging data and the corresponding ground-truth labels. Participants can start designing and training their methods using this training dataset. The validation data will then be released within three weeks after the training data is released. This will allow participants to obtain preliminary results in unseen data and also report these in their submitted short MICCAI LNCS papers, in addition to their cross-validated results on the training data. The ground truth of the validation data will not be provided to the participants, but multiple submissions to the online evaluation platforms will be allowed. The top-ranked participating teams in the validation phase will be invited to prepare their slides for a short oral presentation of their method during the BraTS challenge at MICCAI 2023.

Finally, all participants will be evaluated and ranked on the same unseen testing data, which will not be made available to the participants, after uploading their containerized method in the evaluation platforms. The final top-ranked participating teams will be announced at the 2023 MICCAI Annual Meeting. The top-ranked participating teams of both the tasks will receive monetary prizes. To improve inclusion and participation of teams from underrepresented communities in imaging grand challenges, specifically teams from Africa, where data for the BraTS-Africa challenge was acquired, participating African teams with the best rank will receive Lacuna Equity & Health Prizes. This prize, as subjected to funding, will be limited to teams who completed the BraTS-Africa BrainHack 2023 workshop.

## Acknowledgments

Success of any challenge in the medical domain depends upon the quality of well annotated multi- institutional datasets. We are grateful to all the data contributors, annotators and approvers for their time and efforts.

## Funding

Research reported in this publication was partly supported by the National Institutes of Health (NIH) under award numbers: NIH/NCI/ITCR:U01CA242871. The content of this publication is solely the responsibility of the authors and does not represent the official views of the NIH. This research also received funding in part from Lacuna fund for Machine Learning Datasets for Better Healthcare Outcomes.